\documentclass[twocolumn,11pt]{article}
\usepackage{times}
\usepackage[dvips]{graphicx}
\begin{document}
\global\def\refname{{\normalsize \it References:}}
\baselineskip 12.5pt
%
%
% TITLE, AUTHOR, ABSTRACT, KEYWORDS
%
\title{\LARGE \bf Interleaving Command Sequences: a Threat to Secure Smartcard Interoperability}

\date{}

\author{\hspace*{-10pt}
\begin{minipage}[t]{2in} \normalsize \baselineskip 12.5pt
\centerline{Maurizio Talamo}
\centerline{Nestor Laboratory and}
\centerline{Department of Mathematics}
\centerline{Tor Vergata University of Rome}
\centerline{Italy}
\centerline{talamo@nestor.uniroma2.it}
\end{minipage} \kern 0in
\begin{minipage}[t]{2in} \normalsize \baselineskip 12.5pt
\centerline{Maulahikmah Galinium}
\centerline{Department of Mathematics}
\centerline{Tor Vergata University of Rome}
\centerline{Italy}
\centerline{galinium@nestor.uniroma2.it}
\end{minipage} \kern 0in
\begin{minipage}[t]{2.7in} \normalsize \baselineskip 12.5pt
\centerline{Christian H. Schunck, Franco Arcieri}
\centerline{Nestor Laboratory}
\centerline{Tor Vergata University of Rome}
\centerline{Italy}
\centerline{\{schunck, arcieri\}@nestor.uniroma2.it}
\end{minipage}
%
% If you are three authors then you can use three mini--pages
% instead of two. Their horizontal size must be less than 2.7in
% indicated above. It can be e.g. 2.3in. However, you must pay
% attention that you do not exceed the total width of the text.
%
\\ \\ \hspace*{-10pt}
\begin{minipage}[b]{6.9in} \normalsize
\baselineskip 12.5pt {\it Abstract:}
% The text of the abstract follows.
The increasingly widespread use of smartcards for a variety of sensitive applications, including digital signatures, creates the need to ensure and possibly certify the secure interoperability of these devices. Standard certification criteria, in particular the Common Criteria, define security requirements but do not sufficiently address the problem of interoperability. Here we consider the interoperability problem which arises when various applications interact with different smartcards through a middleware. In such a situation it is possible that a smartcard of type $S$ receives commands that were supposed to be executed on a different smartcard of type $S'$. Such ``external commands'' can interleave with the commands that were supposed to be executed on $S$. We experimentally demonstrate this problem with a Common Criteria certified digital signature process on a commercially available smartcard. Importantly, in some of these cases the digital signature processes terminate without generating an error message or warning to the user.%
\\ [4mm] {\it Key--Words:}
% The key-words follow.
smartcard, common criteria, interoperability, digital signature
\end{minipage}
\vspace{-10pt}}

\maketitle

\thispagestyle{empty} \pagestyle{empty}
% numbers of pages are supplemented by the editor
%
% THE BEGINNING OF THE TEXT
%
\section{Introduction}
\label{introduction} \vspace{-4pt}
Smartcards (SC) are becoming increasingly popular in many countries and are deployed, for example, as credit cards, health cards and electronic identification documents. With these devices users control highly sensitive information and may perform security tasks such as electronic authentication and digital signature. As the importance and world-wide spread of SCs increases, the interoperability of these devices becomes more important along with their security in environments where SCs from different manufacturers and issuers are used at the same time. 

The Common Criteria (CC) \cite{Common_C} and the CWA 14169 \cite{CWA} standards are used to certify the correct behavior of a SC in a well defined environment, i.e. for a specific target of evaluation (TOE). The TOE is precisely described and usually comprises a specific microprocessor, a specified firmware and specified middleware \cite{Common_C}; however, environments where different SCs are used at the same time are usually different from the TOE. In particular, a SC could be confronted with commands from different processes, be it accidentally, on purpose or during an attack. Trusted SC interoperability, therefore, requires a careful analysis of how SCs operate in such situations and the consideration of these results in the design of interoperable systems.

One goal of current research and development efforts regarding SC interoperability is to create a framework that enables the concurrent use of different SCs. These efforts focus on diverse topics such as standardization \cite{ISO7816}, \cite{ISO24727}, \cite{NIST}, architectures for SC based authentication services \cite{Interoperability1}, and open protocols \cite{Interoperability2}.

However, interoperability problems emerge already with the attempt to recognize what type of SC is actually used. In practice, this is currently done by detecting the presence of certain applications on the SC \cite{Interoperability3}. Deducing the SC type from such information is, at its best, an indirect method which might not uniquely identify the SC type, and leave it prone to potential attacks. If the SC type is incorrectly identified, "external commands" will be sent to the SC. Here, we define an "external command" as an Application Protocol Data Unit (APDU) sequence that does not correspond to the regular APDU sequence supplied to the SC in the executable code of the middleware originally used during the security certification (e.g. according to CC). In a setting where different SCs interact with applications via a middleware, APDUs that are supposed to be delivered to a certain SC type $S$ might be received by a SC of type $S'$ (e.g., due to routing errors). In such situations ``external commands'' can interleave with regular commands.

References \cite{RIPSD}, \cite{VECISD} study the behavior of commercial signature SCs during the sequential steps of a digital signature process. First the APDUs sent - in the setting used for the CC certification - from the middleware to the SC were identified. Using a model checking approach, the SCs were then targeted with modified APDUs during the digital signature process of a fixed document. The experiments showed that certain modified commands are accepted by the SCs without errors being generated and demonstrated that CC certification is not sufficient to address the SC interoperability problem.

In this work we address the problem of interleaving commands for SC interoperability by analyzing the situation in which different applications interact with SCs via a middleware. A CC certified digital signature process on a commercially available SC is then tested to demonstrate the relevance of this problem experimentally. Finally, we discuss the complexity of the underlying issues and how the experimental test setup may be improved in the future to identify and prevent potential interoperability problems of this kind.

\section{The Interoperability Problem: Interleaving Command Sequences}
\label{Interoperability} \vspace{-4pt}
To address the interoperability problem on a fundamental level, we consider a straight-line program $P_1$ with steps $S_{1,1}$, $S_{1,2}$, ...., $S_{1,l}$. It is assumed that the straight line program $P_1$ has been certified to produce a correct result if the sequence of commands $C_{1,1}$, $C_{1,2}$, ..., $C_{1,l}$ originally associated with these steps is executed in the correct order and without modifications on a SC of type $S$. For example, the command sequence $C_{1,1}$, $C_{1,2}$, ..., $C_{1,l}$ could match the one supplied to $S$ in the executable code of the middleware that was used for CC certification. We call a command $C_{i,j}$ \emph{``globally legal''} if it is processed in step $S_{i,j}$ of program $P_i$ on SC of type $S$ and the process $P_i$ has been certified on $S$.

\begin{figure}[htb]
\centering
\includegraphics[width=0.8\hsize,clip=true,angle=0]{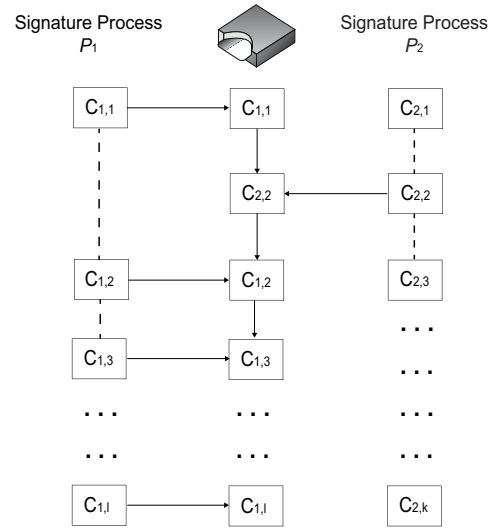}
\caption{Two concurrent signing sessions}
\label{fig:simple}
\end{figure}

In an environment where several applications interact with SCs via a middleware, commands from another straight line program $P_2$ may interleave with the commands from $P_1$ on $S$. Here $P_2$ is a straight line program for another SC type $S'$ with steps $S_{2,1}$, $S_{2,2}$, ...., $S_{2,k}$ and commands $C_{2,1}$, $C_{2,2}$, ..., $C_{2,k}$. Fig.~\ref{fig:simple} illustrates the situation and shows how SC $S$ receives interleaving commands associated with different steps of the two digital signature processes $P_1$ and $P_2$.

We will now analyze this SC interoperability problem in more detail, and in particular we distinguish the following cases:
\begin{enumerate}
\item In step $S_{1,j}$ of $P_1$, SC $S$ receives command $C_{1,j}$ and processes it without error. The digital signature process makes a correct transition to step $S_{1,j+1}$.
\item In step $S_{1,j}$ of $P_1$, SC $S$ receives command $C_{2,i}$ corresponding to step $S_{2,i}$ of $P_2$ and correctly generates an error. At this point the error can be detected and digital signature process can be interrupted.
\item In step $S_{1,j}$ of $P_1$, SC $S$ receives command $C_{2,i}$ corresponding to step $S_{2,i}$ of $P_2$ and processes it without generating an error, i.e. the SC recognizes this command as \emph{``locally legal''}. However, in this case, $C_{2,i}$ is \emph{not} globally legal. We refer to this situation as an \emph{``anomaly''} since it is unknown how the overall signature process will be affected. The program may now potentially make a transition to any step of the two programs $P_1$ or $P_2$.
\end{enumerate}

The first case simply describes the correct process. In the second case, the digital signature process may be interrupted, but the important point is that an error is generated and this error can be detected by the middleware. The interoperability environment can then be designed to handle such situations appropriately. The third possibility, however, poses the real problem for trusted SC interoperability: the ``certified'' and, therefore, trusted process has been modified but no error message has been generated. One anomaly can potentially be followed by several others and finally the digital signature process may terminate with a questionable result. Without receiving an error or a warning, a user cannot know whether all steps in the digital signature process were completed correctly or whether there have been one or more anomalies.

In the following we demonstrate experimentally, with a commercially available CC certified SC for digital signature, how commands from a different straight line program may interleave with the original one. Furthermore we present one example where even though an error is generated, an external command that intersects the original program can render a SC inappropriate for further use. The testing environment developed for this purpose, as well as relevant details about SCs are described in the next section.

\section{The Testing Environment}
\label{testing} \vspace{-4pt}
In our experiments, we study two commercially available SCs from two different manufacturers. The core of a SC is its microprocessor, which contains on board, a cryptographic processor, a small EEPROM random access memory ($\approx$ 64 KBytes), an operating system and a memory mapped file system \cite{SCHandBook}. The microprocessor is customized (masked) in order to execute APDU sent from external software applications through a serial communication line.

The ISO 7816 standard \cite{ISO7816}, specifies the set of APDU that can be implemented by any compatible SC microprocessor. In particular, an APDU consists of a mandatory header of 4 bytes: the Class Byte (cla), the Instruction Byte (ins) and two parameter bytes (p1, p2). The header can be followed by a conditional body of variable length, which is composed by the length (in bytes) of the data field (lc), the data field itself and the maximum number of bytes expected in the data field of the response (le). Responses to any APDU are encoded in a variable length data field and two bytes mandatory return codes.

\begin{figure}[htb]
\centering
\includegraphics[width=1\hsize,clip=true,angle=0]{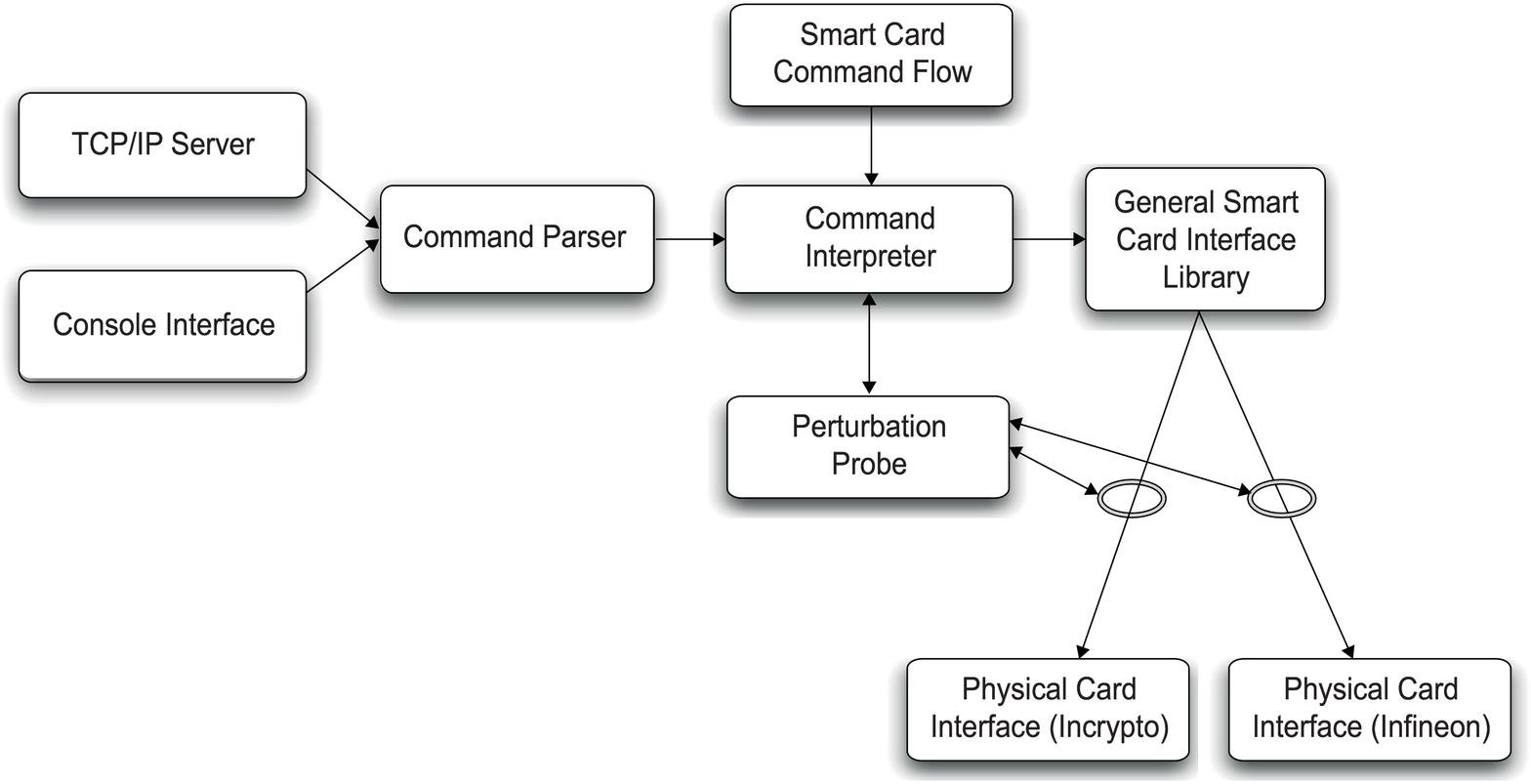}
\caption{Architecture of the Crypto Probing System}
\label{fig:crypto}
\end{figure}

To probe and analyze the SC behavior we have developed a Crypto Probing System (CPS) whose overall architecture is shown in fig.~\ref{fig:crypto}. As each SC uses a different APDU sequence in the digital signature process, the CPS is designed to interface with both SCs used in this project. Effectively, it therefore acts as a \emph{middleware} between the external applications and the real SCs. 

The CPS is able to translate its simplified instructions to the corresponding sequence of APDUs (cla, ins, p1, p2, length and values of the possible annexed data buffer) to be sent to the connected physical SC and to translate the SC responses in a common format. Moreover, to further simplify the interface with the SC, the CPS is given the \emph{globally legal} APDUs to be sent in each step of the digital signature process (SC commands flow), and the CPS is able to generate alternate command sequences to test the SC responses in different situations. This way, the CPS offers a simple interface for testing applications verifying process correctness and robustness on different physical devices and in the presence of interleaving command sequences.

The CPS can be invoked via command line, to interactively test the command sequences, or used as a daemon, which stays in execution and accepts commands on TCP/IP connections. The commands sent via command line are parsed and interpreted by the CPS based on SC library. The elementary instruction of the CPS is made by a single APDU.

\section{Results}
\label{result} 
\begin{figure}[htb]
\centering
\includegraphics[width=0.8\hsize,clip=true,angle=0]{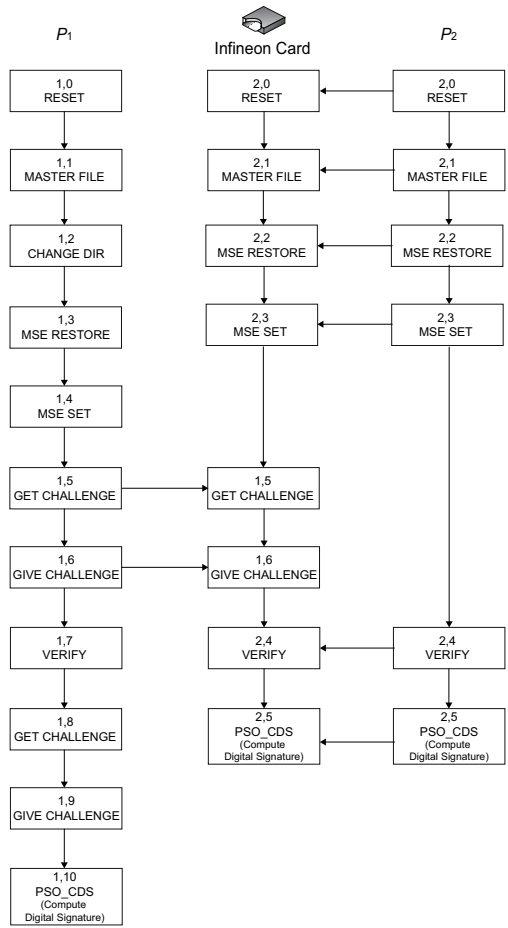}
\caption{Two concurrent signing sessions, Infineon Smartcard}
\label{fig:result1}
\end{figure}

\begin{figure}[htb]
\centering
\includegraphics[width=0.8\hsize,clip=true,angle=0]{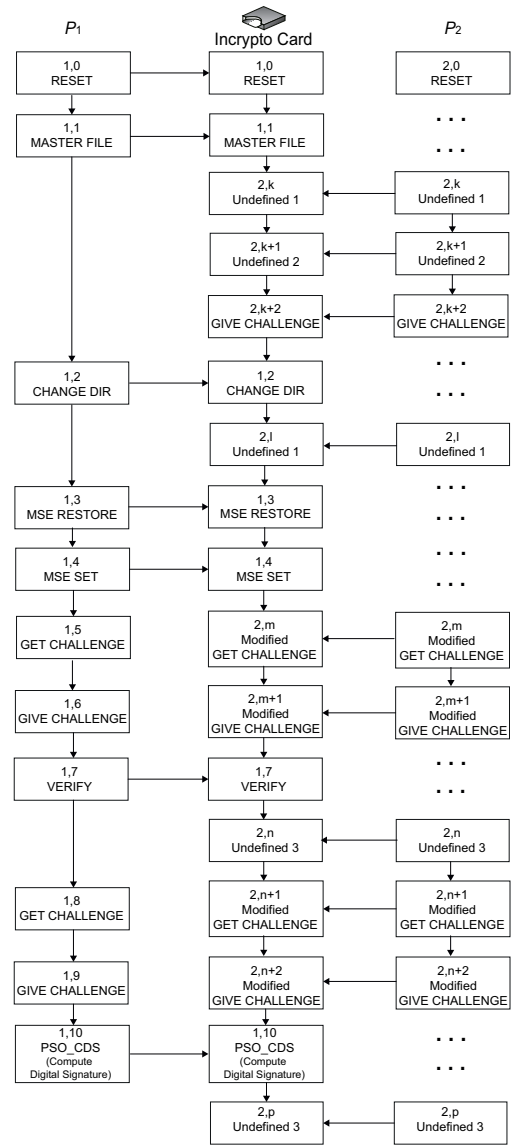}
\caption{Two concurrent signing sessions, Incrypto Smartcard}
\label{fig:result2}
\end{figure}

\begin{figure}
\centering
\includegraphics[width=0.9\hsize,clip=true,angle=0]{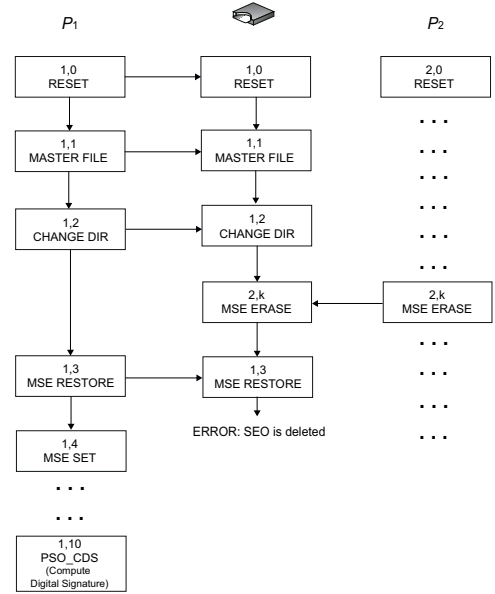}
\caption{Signing session with interleaving ``MSE Erase'' command}
\label{fig:result3}
\end{figure}

\vspace{-4pt}
In this section we present the main experimental results of this work. We use the CPS testing environment to show how external commands interleave with the globally legal commands in a SC based digital signature process. The experiments are carried out with two CC certified SCs from STM-Incrypto34 and Infineon-CardOs \cite{InfineonCC}, \cite{IncryptoCC}. The main results are shown in figs. ~\ref{fig:result1}, ~\ref{fig:result2}, and ~\ref{fig:result3} and tables ~\ref{tab:table1} and ~\ref{tab:table2}. The left (right) column of fig. ~\ref{fig:result1} presents the 10 (5) steps of the digital signatures processes with the Incrypto (Infineon) SCs. Note that we do not count the initial "RESET" and have given similar steps in both processes the same label, although the APDUs associated with these steps may be different.

Fig.~\ref{fig:result1} shows how the commands "Get Challenge" and "Give Challenge" from steps $(1,5)$ and $(1,6)$ from process $P_1$ interleave with steps $(2,1)$ to $(2,5)$ of process $P_2$ on the Infineon SC (central column in fig.~\ref{fig:result1}). No error message is generated, and the process $P_2$ terminates as if no interference occurred. In fact, our experiments show that ``Get Challenge'' and ``Give Challenge'' commands of process $P_1$ can interleave with process P2 before and after all of its steps. In this case, the interleaving commands of two globally legal digital signature processes create a result whose trustworthiness has not been assured. Because a user cannot distinguish this situation from one in which no anomaly occurred, this problem might undermine the overall trustworthiness of SC use in an interoperable environment. Furthermore, sending the ``Get Challenge'' and ``Give Challenge'' commands repeatedly to the SC could be used by an attacker to put a digital signature process effectively on hold.

The results shown in fig.~\ref{fig:result2} were obtained with the Incrypto SC. As above, $P_1$ represents the digital signature process associated with this SC and the globally legal APDUs of each step are given in table~\ref{tab:table1} (here RN is short for ``random number''). Process $P_2$ contains APDUs that are either slightly or substantially different from the globally legal APDUs in $P_1$ (see table~\ref{tab:table2}, the modified parts are printed in bold font). In particular, certain APDUs are not documented for the Incrypto SC: these APDUs are therefore labeled as ``undefined''. The exact sequence of APDUs in $P_2$ is not part of a single digital signature process on any SC we are aware of. Nevertheless, these commands could well be part of such processes implemented on one or several different SCs.

The command sequence executed in our experiments is shown in the central column of fig.~\ref{fig:result2}. Although this executed process contains six additional commands (five of them ``undefined'') and four modified commands, it terminates without any error message. In addition, the sequence can be looped back to the first node (1,1) ``Master File'' after any step of the executed process and afterwards continue until the end. These examples show how drastically digital signature processes can be modified via interleaving commands without the associated anomalies being recognized. An interoperable environment that does not address this issue may not be considered trustworthy and may have vulnerabilities that potential attackers could seek to exploit.

\begin{table}[htb]\footnotesize
\caption{Globally legal APDUs of process $P_1$ in fig.~\ref{fig:result2} (hexadecimal representation)} \label{tab:table1}
	\centering
	\scalebox{1}{
		\begin{tabular}{|l|l|l|l|l|l|l|l|} \hline  Node & \multicolumn{7}{|c|}{Globally legal APDUs}\\\hline
		             &cla &ins & p1 & p2 & Lc &Data  & Le  \\\hline
 1,1    & 00 & A4 & 00 & 00 & 00 & -    & FF       \\\hline
 1,2    & 00 & A4 & 00 & 00 & 02 &14 00  & FF       \\\hline
 1,3    & 00 & 22 & F3 & 03 & 00 & -    & 00       \\\hline
 1,4    & 00 & 22 & F1 & B6 & 03 &83 01 10& 00       \\\hline
 1,5    & 00 & 84 & 00 & 00 & 00 &  -   & 08  \\\hline
 1,6    & 80 & 86 & 00 & 00 & 08 &RN& 00     \\\hline
 1,7    & 0C & 20 & 00 & 9A & 08 &PIN   & 00       \\\hline
 1,8    & 00 & 84 & 00 & 00 & 00 &  -   & 08  \\\hline
 1,9    & 80 & 86 & 00 & 00 & 08 &RN& 00     \\\hline
 1,10   & 0C & 2A & 9E & 9A & 75 &00-74 & FF       \\\hline
		\end{tabular}
	}
\end{table}

\begin{table}[htb]\footnotesize
\caption{Globally legal and modified APDUs of the executed process of fig.~\ref{fig:result2} (hexadecimal representation)} \label{tab:table2}
	\centering
	\scalebox{1}{
		\begin{tabular}{|l|l|l|l|l|l|l|l|} \hline  Node & \multicolumn{7}{|c|}{Globally legal and \textbf{modified} APDUs}\\\hline
		              &cla &ins & p1 & p2 & Lc &Data  & Le    \\\hline
 1,1       & 00 & A4 & 00 & 00 & 00 & -    & FF       \\\hline
 2,k       &\textbf{81}&\textbf{86}&\textbf{00}&\textbf{00}&\textbf{02}&\textbf{14 00}&\textbf{00}\\\hline
 2,k+1     &\textbf{8F}&\textbf{86}&\textbf{00}&\textbf{00}&\textbf{02}&\textbf{14 00}&\textbf{00}\\\hline
 2,k+2     &\textbf{80}&\textbf{86}&\textbf{AC}&\textbf{45}&\textbf{08}&\textbf{RN}&\textbf{00} \\\hline
 1,2       & 00 & A4 & 00 & 00 & 02 &14 00  & FF       \\\hline
 2,l       &\textbf{81}&\textbf{86}&\textbf{00}&\textbf{00}&\textbf{02}&\textbf{14 00}&\textbf{00}\\\hline
 1,3       & 00 & 22 & F3 & 03 & 00 & -    & 00       \\\hline
 1,4       & 00 & 22 & F1 & B6 & 03 &83 01 10& 00       \\\hline
 2,m       & 00 & 84 &\textbf{BD}&\textbf{17}&00&-&08  \\\hline
 2,m+1     & 80 & 86&\textbf{AC}&\textbf{45}&08&RN&00\\\hline
 1,7       & 0C & 20 & 00 & 9A & 08 &PIN   & 00       \\\hline
 2,n       &\textbf{8C}&\textbf{86}&\textbf{00}&\textbf{00}&\textbf{02}&\textbf{14 00}&\textbf{00}\\\hline
 2,n+1     & 00 & 84 &\textbf{BD}&\textbf{17}&00&-&08    \\\hline
 2,n+2     & 80& 86 &\textbf{AC}&\textbf{45}&08&RN&00 \\\hline
 1,10      & 0C & 2A & 9E & 9A & 75 &00-74 & FF       \\\hline
 2,p       &\textbf{8C}&\textbf{86}&\textbf{00}&\textbf{00}&\textbf{02}&\textbf{14 00}&\textbf{00}\\\hline
		\end{tabular}
	}
\end{table}

Finally, we would like to point out a problem caused by interleaving commands that has considerable consequences even though an error is generated. In this experiment (shown in fig.~\ref{fig:result3}), $P_2$ contains the ``MSE Erase'' command. This command is usually not part of a digital signature process as it erases the Security Environment Object (SEO); however, it is conceivable that this command is used by an application interacting with the middleware for some purpose. It may then accidentally, or even in an attack, interleave with a digital signature process like $P_1$. We observe experimentally that ``MSE Erase'', executed as shown in the central column of fig.~\ref{fig:result3}, erases the SEO on the SC without warning, and the digital signature process generates an error after the next step $S_{1,3}$. The digital signature function of the SC is herewith permanently destroyed and a physical replacement of the SC is required. In principal, such vulnerability could be systematically exploited in an attack on all SCs issued by the digital signature service provider because neither PIN nor PUK is required to execute the ``MSE Erase'' command.

\section{Conclusion}
\label{S3} \vspace{-4pt}

The experiments described above show that the problem of interleaving command sequences is serious and that it must be addressed to ensure a secure and trustworthy environment for SC interoperability.

As stated in the introduction, in previous work \cite{RIPSD}, \cite{VECISD} a C-Murphi model checker \cite{CMu} has been used to test SC behavior in the presence of disturbed commands. Model checking can address extended systems which can assume millions of different states \cite{ModCh} and can in principal be used to identify anomalies. However, the complexity of the verification increases exponentially if interleaving commands are to be taken into account: assume that for every step of two digital signature processes, the input command has only a $16$ bits and assume that the two signature processes consist of $10$ steps each. Even under this strong simplification, a brute force model checker may be required to make more than ${20 \choose 10} *2^{16+16}$ tests. This is due to the fact that in this approach all possible sequences that can be obtained by mixing the two signature processes are generated. Note that in an interoperable environment, possibly tens, if not hundreds, of applications may interact concurrently with various SC types via some middleware. As a result, a brute force model checking approach is clearly not a viable solution, especially if it is operated on real SCs as illustrated in the experiments described above where the execution of a single command can take up to $1$ second.

In future research, we plan to extend the model checking approach to avoid brute force testing and to identify errors and anomalies effectively. This can be done if one prevents the model checker from searching through all possible sequences of anomalies and errors by taking the results of the already existing CC certification into account. Such an efficient model checker can then be integrated into a middleware as a ``watch-dog'' to identify an anomaly as it occurs and to prevent computational chains with two or more anomalies. In this case, it will be possible to extend the CC to certify the anomaly-free interoperability of several SC applications interacting via a middleware with different SC types.


\begin{thebibliography}{11}

\vspace{-7pt}
\bibitem{Common_C}
{\em Common Criteria for Information Technology Security Evaluation, version 3.1}, Common Criteria Std. CCMB-2009-07-001, CCMB-2009-07-002, CCMB-2009-07-003, Rev. 3 FINAL, 2009.

\vspace{-7pt}
\bibitem{CWA}
{\em Secure signature-creation devices \emph{``EAL 4+''}}, European Committe for Standardization (CEN) Std. CWA 14169, Mar. 2004.

\vspace{-7pt}
\bibitem{ISO7816}
{\em Identification cards -— Integrated circuit cards -— Part 4: Organization, security and commands for interchange},  International Organization for Standardization Std. ISO/IEC 7816-4:2005, Jan. 2005.

\vspace{-7pt}
\bibitem{ISO24727}
{\em Identification cards -– Integrated circuit cards programming interfaces -– Part 3: Application programming interface},  International Organization for Standardization Std. ISO/IEC 24727-3:2008, Dec. 2008.

\vspace{-7pt}
\bibitem{NIST}
T.~Scwarzhoff, et al, {\em Government Smart Card Interoperability Specification, version 2.1}, United States of America National Institute for Standards and Technology (NIST), Tech. Rep. 6887, 2003.

\vspace{-7pt}
\bibitem{Interoperability1}
A.~Tauber, et al, Towards Interoperability: An Architecture for PAN-European eID-Based Authentication Services, in {\em EGOVIS 2010}, 2010, pp.~120-133.

\vspace{-7pt}
\bibitem{Interoperability2}
T.~Cucinotta, M.~Di Natale and D.~Corcoran, A Protocol for Programmable Smart Cards, in {\em Proceedings of the 14th International Workshop on Database and Expert Systems Applications (DEXA'03)}, IEEE Computer Society, 2003

\vspace{-7pt}
\bibitem{Interoperability3}
D. H\"uehnlein and M. Bach, How to Use ISO/IEC 24727-3 with Arbitrary Smart Cards, in {\em TrustBus 2007}, ser. LNCS vol. 4657. Springer Verlag, 2007, pp. 280-289.

\vspace{-7pt}
\bibitem{RIPSD}
M.~Talamo, et al, Robustness and Interoperability Problems in Security Devices, in {\em Proceedings of 4th International Conferences on Information Security and Cryptology, INSCRYPT 2008}, 2008.

\vspace{-7pt}
\bibitem{VECISD}
M.~Talamo, et al, Verifying extended criteria for the interoperability of security devices, in {\em Proceedings of 3rd International Symposium on Information Security, IS08}, ser. LNCS vol. 5332. Springer, 2008, pp. 1131-1139.

\vspace{-7pt}
\bibitem{SCHandBook}
W.~Rankl and W.~Effing, {\em Smart Card Handbook}, 4th ed. West Sussex, UK: Wiley, 2010.

\vspace{-7pt}
\bibitem{InfineonCC}
Siemens, {\em Certification Report CardOS V4.2 CNS with Application for Digital Signature}.

\vspace{-7pt}
\bibitem{IncryptoCC}
{\em Secure Signature Creation Device Incrypto34v2 from ST INCARD S.r.l}, {Bundesamt} f\"ur Sicherheit in der Informationtechnik Std. BSI-DSZ-CC-0202-2005, 2005.

\vspace{-7pt}
\bibitem{CMu}
G. Della Penna. (2005, Aug) The CMurphi Verifier. [Online]. Available: http://www.di.univaq.it/gdellape/murphi/cmurphi.php.

\vspace{-7pt}
\bibitem{ModCh}
E. M.~Clarke, O.~Grumberg and D. A.~Peled, {\em Model Checking}. The MIT Press, 1999.


\end{thebibliography}
\end{document}